# Highly reliable, ultra-wideband, isolator-free quantum-dot mode-locked frequency combs for optical interconnects beyond 3.2Tb/s

Shujie Pan[1,2], Victoria Cao[3], Yiheng Feng[1], Dingyi Wu[4], Jie Yan[4], Junjie Yang[2], Chao Zhao[1,5], Xi Xiao[4*], and Siming Chen[1,5*]

[1]*Laboratory of Solid State Optoelectronics Information Technology, Institute of Semiconductors, Chinese Academy of Sciences, Beijing 100083, China*
[2]*HS Photonics Co., Ltd., Xiangjiang Science & Technology Innovation Base, Changsha, Hunan 413000, China*
[3]*College of Integrated Circuits and Optoelectronic Chips, Shenzhen Technology University, 3002 Lantian Road, Shenzhen 518118, China*
[4]*National Information Optoelectronics Innovation Center, China Information and Communication Technologies Group Corporation (CICT), Wuhan 430074, China*
[5]*College of Materials Science and Opto-Electronic Technology, University of Chinese Academy of Science, Beijing 101804, China*

Correspondence: Siming Chen (smchen@semi.ac.cn); Xi Xiao (xiaoxi@noeic.com)

**Abstract**

Quantum dot mode-locked laser-based optical frequency combs are emerging as a critical solution for achieving low-cost, high-efficiency, and large-capacity optical interconnects. The practical implementation of wavelength division multiplexing interconnects necessitates a temperature-stable OFC source with a minimum 100 GHz channel spacing to enable high-bandwidth modulation while mitigating the complexity of optical filtering and detection. By leveraging the advanced co-doping technique and a colliding pulse mode-locking scheme, here, we report a compact, ultra-wideband, highly reliable, isolator-free 100 GHz-spacing InAs/GaAs QD OFC source operating up to a record temperature of 140 °C. The comb source delivers a record 3 dB optical bandwidth of 14.312 nm, containing 26 flat-top comb lines, each supporting 128 Gb/s PAM-4 modulation, which results in a total throughput of 3.328 Tb/s with an extremely low power consumption of 0.394 pJ/bit at 25°C. Performance remains stable at 85 °C, with negligible degradation of device critical metrics. Remarkably, accelerated aging tests under harsh conditions (85 °C with 8× threshold current injection) revealed a mean time to failure of approximately 207 years. The QD OFC source demonstrated in this work, for the first time, establishes a concrete link between fundamental research on comb sources and their practical deployment in next-generation, high-density optical interconnect systems.

**Introduction**

Artificial intelligence (AI) and high-performance computing (HPC) are driving revolutionary advances in optical input/output (OIO) technologies through their unprecedented bandwidth demands. These advances particularly focus on wavelength division multiplexing (WDM) architecture, which relies on highly efficient multi-wavelength light sources to meet performance requirements[1]. Optical frequency combs (OFCs) have emerged as a transformative solution, with their phase-locked spectral lines enabling parallel data transmission while overcoming the power consumption and scalability limitations of discrete laser arrays. Emerging breakthroughs in OFCs have shown remarkable potential to address the escalating requirements for high-speed, high-capacity, and energy-efficient data transmission in next-generation communication systems[2].

Concurrently, semiconductor quantum dot (QD) materials have emerged as a promising platform for OFC generation, leveraging their intrinsic advantages such as superior temperature stability[3-5], low threshold current[5,6], ultra-broad gain spectrum[7] and ultra-fast carrier dynamics[8,9]. Significant progress has been reported in the use of QD-MLL OFCs for high-speed transmission. For example, 32 Gbaud-per-channel Nyquist PAM-4 modulation was demonstrated using a 20 GHz QD-MLL with a 6.1 nm bandwidth[10], and an aggregate data rate of 12.1 Tbit/s was achieved using a 58.2 GHz QD-MLL featuring

26 channels, each modulated with 32 Gbaud QAM[11]. Despite that, channel spacing of at least 100 GHz is preferred to enable high-bandwidth modulation and simplify optical filtering and detection as advocated by the Continuous-Wave Wavelength-Division-Multiplexed (CW-WDM) standard[12]. In the most recent reported studies on a 100 GHz QD comb laser, Chen et al. demonstrated a comb source capable of 8×100 Gb/s/λ PAM-4 data transmission with a laser power consumption of 1.24 pJ/bit[13]. Meanwhile, Rauter et al. reported a record 24-channel comb source, with each channel supporting up to 106 Gb/s/λ PAM-4 modulation using a free-space optical isolator, achieving an aggregate transmission capacity of 2.544 Tb/s[14]. However, all reported results were obtained at room temperature. The practical deployment of OFC-based OIO systems faces significant challenges, particularly in ensuring robust operational stability and maintaining broad optical bandwidth under elevated temperatures. For real-world applications, an ideal OFC source must meet stringent requirements, including sufficient operational lifetime and consistent performance across the full industrial temperature range of -40 °C to 85 °C[15,16]. However, to date, the high-temperature transmission capabilities of 100 GHz QD comb lasers have yet to be demonstrated.

To enhance the high-temperature performance of QD devices, p-type doping has been a traditional focus in previous studies, with a common strategy involving partial modulation doping of GaAs barriers with beryllium[17-19]. However, this approach often leads to an increased threshold current, which can negatively impact device performance to some extent[20,21]. On the other hand, direct n-type doping with silicon in the active region can effectively reduce the threshold current, but typically at the expense of high-temperature stability[22]. The emergence of co-doping techniques, combining the benefits of both n-type and p-type doping, offers a promising pathway to optimize QD device performance in simultaneously lowering the threshold current, increasing the output power, and enhancing high-temperature stability[23-26]. These advancements make co-doped QD devices highly suitable for generating robust OFCs with broad bandwidth, even under demanding operating conditions.

In this work, we demonstrated a significant breakthrough in QD-MLL technology by implementing a co-doping strategy during QD material growth. A colliding-pulse mode-locked (CPML) configuration was utilized for achieving a high repetition rate of 100 GHz, which is desired by OIO applications. Comprehensive device characterization and mode-locking behaviors were investigated across a wide temperature range, from 25 °C to 140 °C. High-speed transmission experiments conducted at both 25 °C and 85 °C serve as a critical method of evaluating the reliability of isolator-free QD-MLL comb lasers under thermal varying conditions. The corresponding laser power consumption per transmitted bit was also quantified. Furthermore, the longevity of the QD-MLL was assessed using the mean time to failure (MTTF) method, based on the evolution of threshold current measured over 1500 hours at 85 °C under a constant stress current of 145 mA (8× threshold current injection). These evaluations provide valuable insights into the impact of temperature fluctuations on device performance, enabling a robust assessment of thermal stability and long-term reliability. In particular, the findings provide critical guidance for deploying QD-MLL OFCs in thermally challenging environments, which is essential for enabling energy-efficient, reliable, and high-density optical interconnects in next-generation intelligent computing networks, including data centers and AI infrastructure.

**Results**

Figure 1a schematically presents the epitaxial structure of the QD comb laser, which was grown on an N-type GaAs (001) substrate by solid-source molecular beam epitaxy (MBE). The details can be found in the Materials and Methods section. In this work, a co-doping strategy was implemented during QD growth to leverage the advantages of both n-type and p-type doping, effectively reducing the threshold current and enhancing optical output power while preserving exceptionally high-temperature stability[23,24,27]. The cross-sectional scanning transmission electron microscopy (TEM) image of the QD active region (Figure 1b) distinctly reveals a visible defect-free dot-in-well (DWELL) active region. A single QD exhibits ~26 nm in diameter and ~7 nm in height, as shown in the inset of Figure 1b. The atomic force microscopy (AFM) image for an uncapped QD sample with the same growth condition is shown in the inset of Figure 1c, where a high dot density of $5.5 \times 10^{10}$ cm$^{-2}$ is obtained. The room-temperature photoluminescence (PL) emission spectrum of the as-grown sample exhibits a prominent peak at approximately 1281.4 nm with a full-width at half-maximum (FWHM) of 31.2 meV, indicating a high degree of size uniformity present among the QDs. It should be mentioned that the FWHM obtained in this work is slightly broader than that reported in previous results[28]. This intentional broadening aims to achieve a broader gain spectrum, thereby accommodating a larger number of comb lines within the 3 dB bandwidth while still maintaining a relatively high peak gain. The quantized energy separation between the ground state (GS) and first excited state (ES1) (1178.8 nm) reaches 84.2 meV, a critical feature for suppressing the carrier escaping from GS to higher energy levels at elevated temperatures[29].

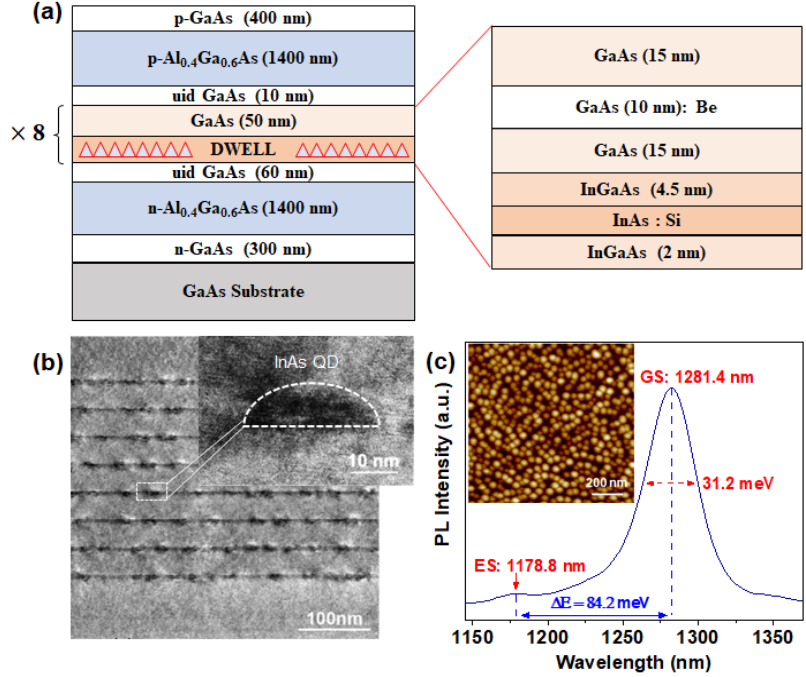

*Figure 1. Epitaxial structure and material property of the QD material. (a) Schematic of the layer structure. (b) Cross-sectional TEM image of the QD active region. (c) RT PL spectrum of QD material. Inset: the AFM image of the uncapped QD sample within $1\mu m^2$ area.*

The fabricated 100 GHz QD-MLL-based comb laser was butterfly-packaged for subsequent testing, as shown in Figure 2a. Notably, conventional MLL-OFCs typically require optical isolators during packaging or testing to protect laser diodes from reflected light and suppress the injection noise caused by back-reflected emission into the laser cavity[30]. However, the use of optical isolators increases system complexity, cost, and poses significant challenges for integration. In contrast, our QD-based implementation capitalizes on its inherent optical feedback resilience, offers a distinct advantage by eliminating the need for optical isolators throughout the entire process, from device packaging to experimental characterization. This advancement significantly reduces system size, weight, power, and cost (SWaP-C), thereby enhancing the practicality and commercial potential of our 100 GHz QD comb lasers for real-world deployment.

In our prior research, a fundamental mode configuration was typically utilized to generate OFCs with a 100 GHz repetition rate. This was achieved using a conventional two-section mode-locked laser (MLL) with a cavity length of 405 μm at an effective index of 3.7[31,32]. However, in this study, a second-order CPML design was adopted following a comprehensive evaluation of the maximum output power of individual comb lines and the 3 dB bandwidth of the laser devices. This approach enhances the optical gain by incorporating a longer cavity length while preserving the high repetition rate. As illustrated in Figure 2b, the total length of the laser cavity is 810 μm, double the length of a traditional two-section MLL operating at the same repetition rate. The SA section is located at the midpoint of the entire laser cavity, with two symmetrically distributed gain sections on either side. This configuration enables the formation of two colliding pulses within the cavity, resulting in second-order harmonic pulses at a 100 GHz repetition rate, effectively doubling the original fundamental frequency of 50 GHz. The cross-sectional scanning electron microscopy (SEM) image of the fabricated device is shown in Figure 2c, depicting a ridge width of 1.69 µm, which is designed to guarantee single-transverse mode operation and to optimize optical coupling efficiency.

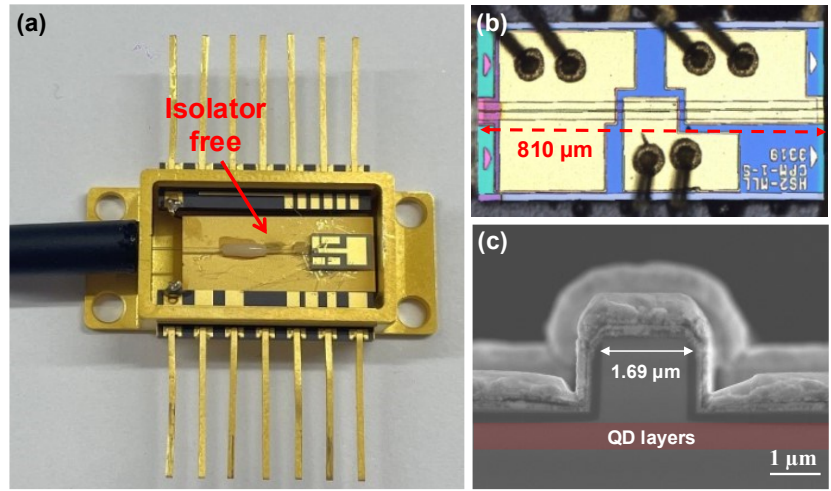

*Figure 2. (a) Photograph of the isolator-free 100 GHz QD comb laser after wire-bonding and butterfly packaging. (b)The top-view microscopic image of the 100 GHz QD comb laser. (c) Cross-sectional SEM image of the fabricated QD comb laser.*

**Device characterization at 25 °C**

Figure 3 shows the characterization of the fabricated 100 GHz QD comb laser at room temperature (25 °C) under continuous wave (CW) conditions. The typical light-current-voltage (L-I-V) performance at various reverse bias voltages is displayed in Figure 3a. As observed, the threshold current increases from 14 mA to 20 mA with increased reverse bias voltage due to the enhanced absorption loss in the SA region. Consequently, the slope efficiency drops from 0.385 W/A to 0.315 W/A, which is consistent with theoretical expectations[33]. To investigate the mode-locking characteristics of the QD comb laser, optical spectral mapping was conducted by sweeping the injection current from 0 mA to 150 mA and the SA reverse bias voltage from 0 to -5 V. The color scheme in Figure 3b represents the number of comb lines within the 3 dB spectral bandwidth of the comb laser. A broad mode-locking region was obtained, and the number of comb lines within the 3dB bandwidth exhibits a positive correlation with increasing injection currents and higher SA reverse bias voltages[34]. The operating conditions yielding the maximum number of comb lines were selected for the subsequent experiments, achieving 26 comb lines at an injection current of 145 mA and a reverse bias voltage of -4.5 V. The corresponding OSA spectrum, presented in Figure 3c, exhibits an ultra-flat comb spectrum with an extensive 3 dB bandwidth of 14.312 nm, measured at a resolution of 0.02 nm. A minimum OSNR of 35 dB is observed for all tones within the 3 dB bandwidth. The high-resolution OSA spectrum measured with a resolution of 0.04 pm, is shown in Figure 3d. As can be observed, the mode spacing for adjacent comb lines is 0.565 nm, corresponding to a repetition rate of 100.65 GHz. To the best of our knowledge, this represents a 100 GHz mode-locked OFC source, featuring the widest 3 dB bandwidth (14.312 nm) ever reported and the most significant number of comb lines (26 tones) to date. The corresponding autocorrelation trace measured at this working point is displayed in Figure 3e, which conforms well to the fitted Gaussian pulse profile. Deconvolution analysis reveals a sub-picosecond pulse duration, achieved without the use of any external pulse compression scheme. To simplify the temperature-dependent experiments and minimize potential confusion, the comb lines initially labelled as tones 6, 9, 12, 15, 18, and 21 in Figure 3c (located within the 3dB bandwidth at 25 °C) were renumbered as Tone 1 through Tone 6 and subsequently tracked across various operating temperatures.

To gain insight into the optical signal quality and evaluate the practical applicability of our 100 GHz QD comb laser in high-speed optical communications, we characterized the relative intensity noise (RIN) performance of both individual comb lines and the entire optical spectrum. As illustrated in Figure 3f, the noise performance exhibits remarkably consistent performance across different tones, reflecting the exceptional uniformity of the comb lines. The integrated average RIN values for the filtered tones and the entire spectrum are approximately -130 dB/Hz and -135 dB/Hz, respectively, over the 0–100 MHz range. In the frequency range from 100 MHz to 20 GHz, these values reach -150 dB/Hz for the filtered tones and -155 dB/Hz for the entire spectrum. Previous studies present at least 20 dB/Hz degradation in RIN values of individual comb lines compared to the entire spectrum[32,35,36]. This discrepancy is attributed to laser mode partition noise, which significantly affects individual filtered comb lines but becomes negligible when considering the entire spectrum, as the intensity fluctuations among the numerous comb lines tend to average out and cancel each other[37-39]. However, our results reveal an ultra-small difference between the RIN values of individual comb lines and those of the entire spectrum, indicating the excellent noise uniformity and stability of the comb source. Moreover, these RIN characteristics meet the specifications outlined in the CW-WDM MSA[12] and the OIF-ITLA-MSA[40], validating the suitability of our comb laser for advanced modulation formats such as PAM-4 in high-speed transmission systems.

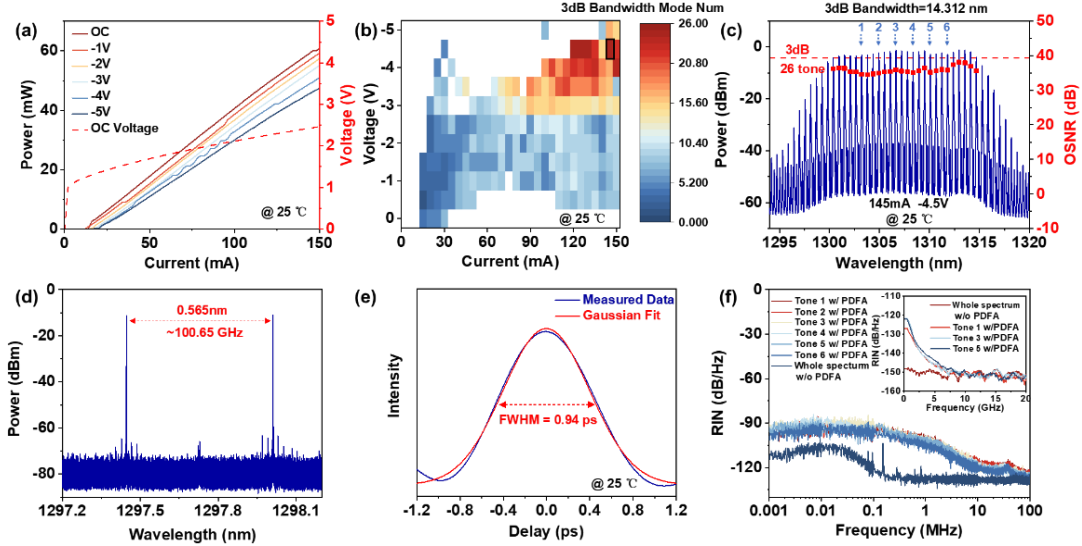

*Figure 3 Characterization of the 100 GHz comb laser at 25 ℃. (a) L-I-V performance of the QD comb laser. (b) Number of tones within 3 dB bandwidth under varied injection current and reversed bias voltage. (c, d) Optical spectra measured with resolutions of 0.02 nm and 0.04 pm, respectively, at an injection current of 145 mA and a reverse bias voltage of -4.5 V. (e) Corresponding autocorrelation trace with Gaussian pulse fitting. (f) RIN of filtered tones and the whole spectrum.*

**Comparison of device characterization at 25 °C and 85 °C**

To assess the functional integrity and the thermal reliability of the 100 GHz comb laser in harsh environments, we systematically evaluated its performance at an elevated temperature of 85 °C and compared the outcomes with those obtained at 25 °C. Figure 4a shows the L-I characteristics and wall-plug efficiency (WPE) curves of the 100 GHz QD comb laser measured at 25 °C and 85 °C, respectively. As seen, the device performance at 85 °C remains comparable to that at 25 °C after packaging. It is also worth noting that the non-optimized, home-fabricated packaging method introduced a coupling loss exceeding 3 dB. This is evidenced by the significant reduction in output power (from 60 mW for the CoC at 150 mA to 26.82 mW for the packaged device) and a corresponding drop in the maximum WPE (from 20.98% to 10.97%). Since all experiments in this work were conducted after device packaging, the actual performance of our comb laser is anticipated to surpass the reported values. Figure 4b presents the optical spectral mapping measured at 85 °C. Although the mode-locking range is slightly narrower than the 25 °C results, it still spans a relatively broad range. A maximum of 22 comb lines within the 3dB bandwidth was achieved at an injection current of 145 mA and a reverse bias voltage of -3.5 V, only 4 fewer than those observed at 25 °C. Moreover, as shown in Figure 4c, the OSNR of each comb line remains above 35 dB. Tone 1 through Tone 6, as previously defined in Figure 3c, corresponds to tones 4, 7, 10, 13, 16, and 19 within the 3 dB bandwidth of this spectrum. The variation in mode spacing over the temperature range from 25 °C to 85 °C is illustrated in Figure 4d. As shown, mode spacing remains remarkably stable across temperatures, with fluctuations limited to 0.07 GHz, which is highly desirable for optical communication systems. This observation aligns well with our previously reported findings and reflects a well-balanced interplay between thermal expansion and refractive index variation [28]. The pulse spectrum at this operating point is presented in Figure 4e, further confirming a stable mode-locked state of the device. In addition, the RIN performance in the range of 0-100 MHz at 85 °C closely matches that at 25 °C (Figure 4f), further validating the temperature-insensitive nature of our QD comb lasers, indicating their potential for TEC-free operation and highlighting their suitability for practical deployment across a wide operating temperature range.

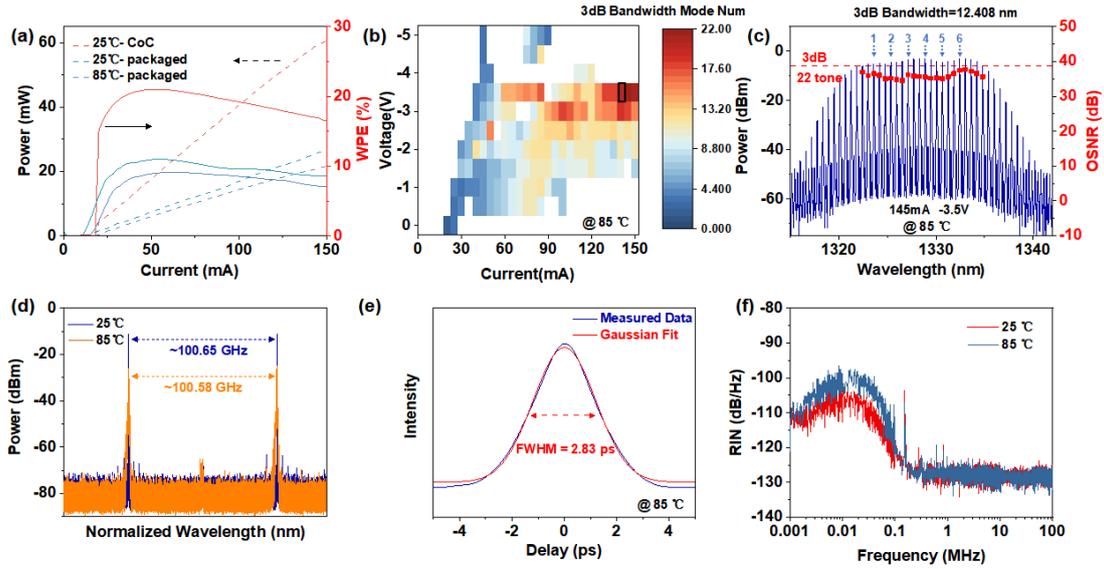

*Figure 4. Characterization comparison of the 100 GHz QD comb laser at 25 °C and 85 °C. (a) L-I characteristics and WPE of the 100 GHz QD comb laser. (b) Number of comb lines within the 3 dB bandwidth under varied injection currents and reverse bias voltages at 85 °C. (c, d) Optical spectra measured with resolutions of 0.02 nm and 0.04 pm, respectively, under an injection current of 145 mA and a reverse bias voltage of -3.5 V at 85 °C. (e) Corresponding autocorrelation trace with a Gaussian pulse fit. (f) RIN of the entire comb spectrum.*

**High-temperature Device Characterization**

A series of high-temperature tests, including L-I characterization and OSA mapping, were further conducted to evaluate the operational temperature limits of the QD comb laser. Figure 5a shows the temperature dependent L-I curves at reverse bias voltage of 0 V. CW lasing was sustained without significant thermal rollover behavior up to an operating temperature of 140 °C. At each temperature, the optimal operating point yielding the maximum number of comb lines within a 3 dB bandwidth was identified via OSA mapping, by which the corresponding OSA spectra was recorded. The center wavelength of these spectra as a function of temperature is plotted in Figure 5b, where a continuous redshift at a rate of approximately 0.49 nm/°C can be observed. This shift is primarily attributed to a combination of bandgap shrinkage and the temperature-dependent shift of the bandgap[41]. The temperature sensitivity of the lasing wavelength could potentially be minimized by precisely engineering the QD structure[42,43]. Figures 5c and 5e present the OSA mapping at 125 °C and 140 °C, respectively. The optimal operating points at these temperatures are marked with the corresponding spectra shown in Figures 5d and 5f. Notably, at 125 °C, up to 9 comb lines remain within the 3 dB bandwidth. Although the overall mode-locking range is significantly reduced at 140 °C due to increased cavity losses, a stable mode-locked state is still sustained, with 5 comb lines retained within the 3 dB bandwidth, each exhibiting an OSNR exceeding 25 dB. For the first time, we demonstrated stable mode-locking at temperatures up to 140 °C, and even at 125 °C, a broad mode-locking range with up to 9 comb lines within the 3 dB bandwidth was retained.

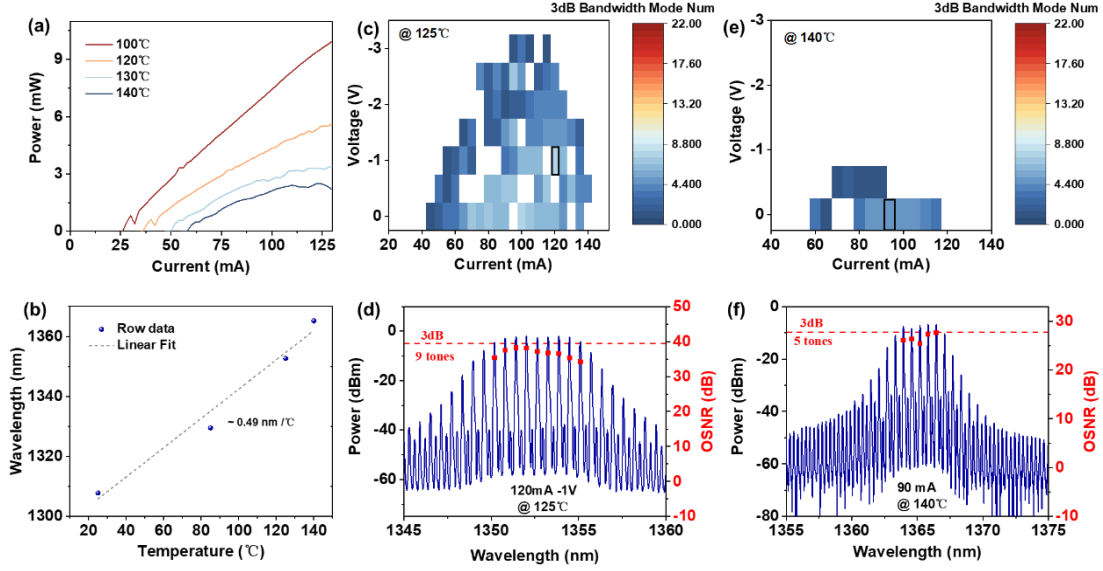

*Figure 5 High-temperature characterization of the 100 GHz QD comb laser. (a) L-I curves of the QD comb laser measured at elevated temperatures with reverse bias voltage of 0 V. (b) Centre wavelength as a function of temperature. (c, e) Number of comb lines within the 3 dB bandwidth under varied injection currents and reverse bias voltages at 125 °C and 140 °C, respectively. (d, f) Optical spectra showing the maximum number of comb lines within the 3 dB bandwidth at 125 °C and 140 °C, respectively.*

**High-speed transmission**

To further evaluate the reliability of our device under practical operating conditions, high-speed transmission experiments were carried out at both 25 °C and an elevated temperature of 85 °C. The transmission setup is illustrated in Figure 6a, which adopts a configuration similar to that reported in Ref 32, with the key distinction being the integration of a sampling oscilloscope for capturing the optical eye diagrams. It should be re-emphasized that no optical isolators were used throughout the experiments, further demonstrating the robustness of our QD comb laser against optical feedback. The external modulation capability of this 100 GHz mode-locked comb laser was evaluated through assessment of the optical eye diagram performance, which is presented in Figure 6b. Clear eye openings were observed for all selected tones at both 25 °C and 85 °C, demonstrating good signal integrity across the temperature range. The transmitter and dispersion eye closure quaternary (TDECQ) values for all tones remain remarkably low under 64 Gbaud PAM-4 (128 Gbit/s) operation, ranging from 1.44 dB to 2.06 dB at 25 °C, and from 2.33 dB to 3.72 dB at 85 °C. The TDECQ penalty across the temperature range is less than 1.8 dB, indicating minimal signal degradation and demonstrating the excellent temperature robustness of the QD comb laser[44]. The measured BER versus optical power of each tone at various temperatures are plotted in Figure 6c & Figure 6d. The hard-decision forward error correction (HD-FEC) threshold at 3.8E-3 (7% overhead) and the soft-decision forward error correction (SD-FEC) threshold at 2.2E-2 (20% overhead) have been labelled. The line-to-line variation of the comb laser is small, as witnessed by the similar slope efficiency and the overlapping points shown in the graph. The receiver sensitivity of this comb laser (at BER = 3.8E-3) equals -5 dBm at 25 °C, and the power penalties between the two temperatures are about 1.5 dB. To the best of our knowledge, this work demonstrates the first high-temperature operation (85 °C) for high-speed transmission (64 Gbaud PAM-4) based on a 100 GHz QD-MLL comb source.

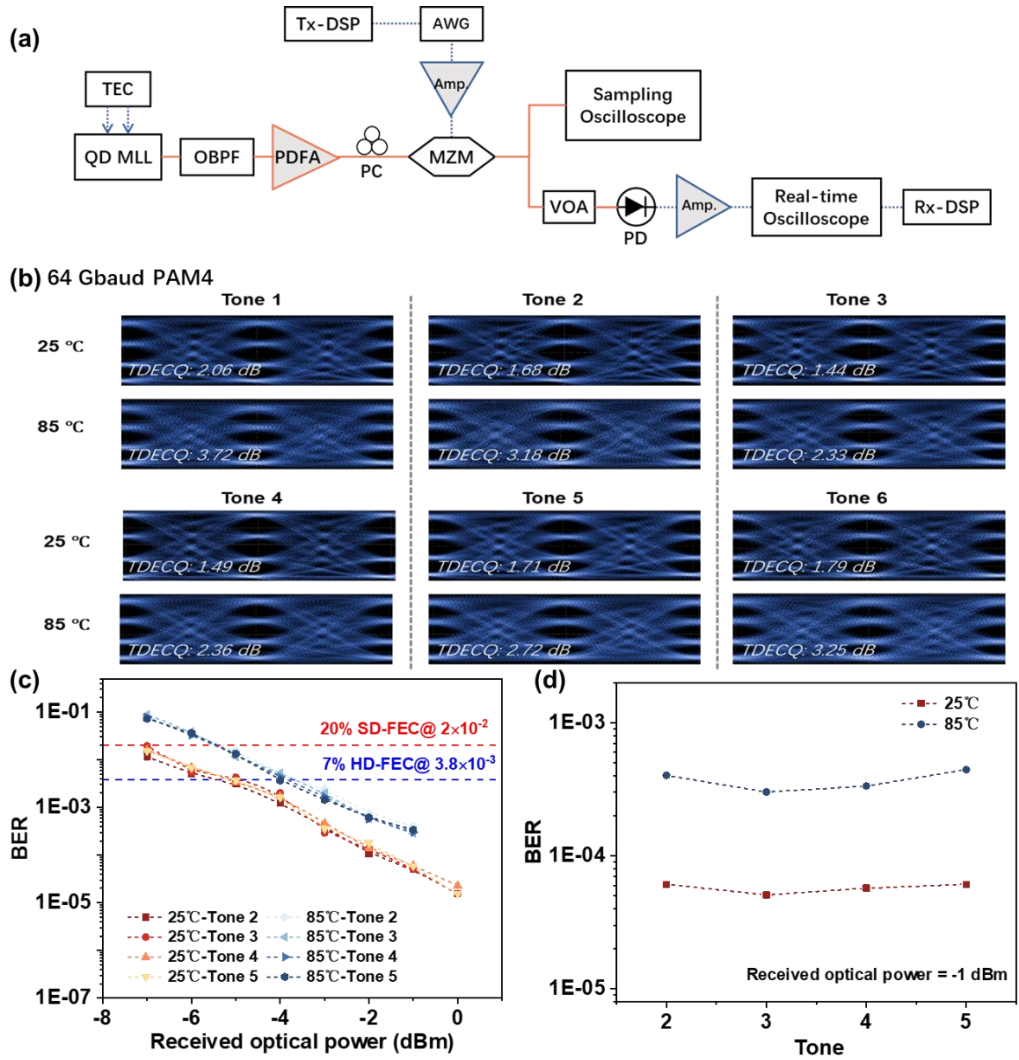

*Figure 6 (a) Data transmission setup, including DSP, digital signal processing; AWG, arbitrary waveform generator; Amp, RF amplifier; MZM, Mach-Zehnder modulator; OBPF, optical band pass filter; PC, polarization controller; VOA, variable optical attenuator; PD, photodetector. (b) 64 Gbaud PAM-4 optical eye diagrams of Tone 1 through Tone 6 at 25 °C (Top) and 85 °C (Bottom). (c) Measured BER as a function of receiving optical power for B2B transmission. (d) BER performance of filtered tones at a received optical power of -1 dBm.*

## Lifetime Estimation

To further validate the long-term reliability of the device, accelerated aging experiments were conducted, and its operational lifetime was estimated based on a mean-time-to-failure (MTTF) analysis, where MTTF is defined as the time required for the threshold current to double its initial value. For this purpose, the 100 GHz QD comb laser was aged over 1500 hours at an ambient temperature of 85 °C, under a constant DC injection current of 145 mA (8× threshold current injection). The evolution of the threshold current as a function of operating time is plotted in Figure 7a. By carefully fitting the age data, the MTTF is estimated to be 1,813,320 hours, corresponding to more than 207 years of operational lifetime. Figure 3b presents the variation in threshold current as a percentage, offering a clear visual representation of its correlation with aging time. The extrapolated intersection of the fitted trend line with the upper horizontal axis indicates an estimated device lifetime of approximately 209 years, aligning well with the calculated MTTF.

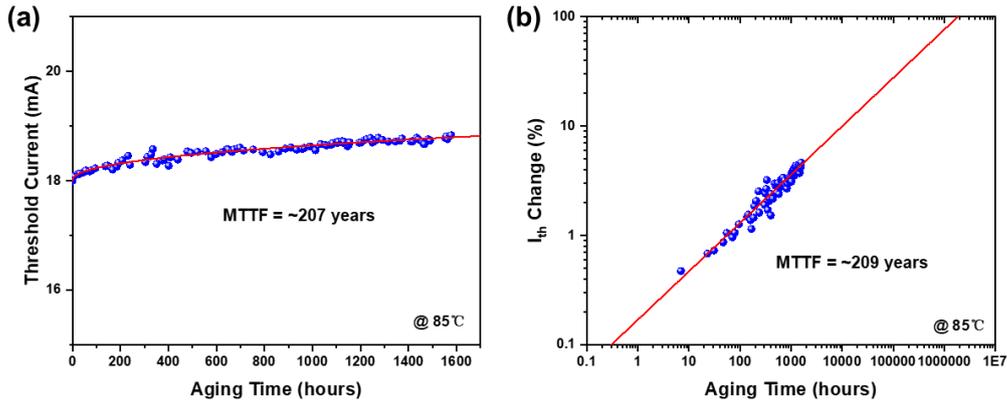

*Figure 7 Threshold current variation with aging time at 85 °C. Axes are in (a) linear form and (b) exponential form, respectively.*

## Discussion

QD-MLL-OFCs have emerged as promising multi-wavelength light sources for optical interconnects and are garnering growing research interest. For future commercialization, particularly in scenarios requiring low cost, low power consumption, and high-density integration, factors beyond key performance metrics such as threshold current, output power, and repetition rate must be considered. Practical deployment necessitates additional capabilities, including wide-temperature-range operation without TEC, resilience to optical feedback to eliminate the need for optical isolators, and overall suitability for miniaturized, cost-effective, and highly integrated systems. Most crucially, long-term reliability under harsh environmental conditions is a fundamental prerequisite for any viable light source intended for real-world applications. Table 1 summarizes recent advances in 100 GHz QD-MLL comb lasers, evaluated against the aforementioned key performance metrics. Among the 100 GHz-spacing comb lasers surveyed, the device reported in this work demonstrates state-of-the-art performance in terms of 3 dB optical bandwidth, transmission capacity, high-temperature operation, and long-term reliability. Notably, all these results were accomplished without the use of an optical isolator.

**Table 1 Comparison of recent 100 GHz QD-MLL comb laser on various material platforms and structures**

| Year/Ref | Material | Mode-locking Method | Repetition Rate (GHz) | 3 dB Bandwidth (nm) | Channel x Bandwidth (Gb/s) | Operating Temperature (°C) | Isolator | Lifetime (Year) |
|---|---|---|---|---|---|---|---|---|
| 2006/[45] | InAs/GaAs QD | CPM-3rd harmonic | 118 | ~0.8 | N/A | 15 | N/A | N/A |
| | | CPM-6th harmonic | 237 | ~1.5 | N/A | 15 | N/A | N/A |
| 2011/[46] | InAs/GaAs QD | Fabry-Perot Etalon | 100 | NA | N/A | | w | N/A |
| 2018/[47] | InAs/GaAs QD | CPM-6th harmonic | 102 | 6.5 | 12 x 10 Gb/s NRZ | 25 | NA | N/A |
| 2019/[48] | Chirped QD | CPML-5th harmonic | 100 | NA | 8 x 112 Gb/s PAM-4 | 18 | NA | N/A |
| 2022/[32] | InAs/GaAs QD | Two-Section | 94 | 3.18 | 7 x 128 Gb/s PAM-4 | 20 | w | N/A |
| 2022/[35] | InAs/GaAs QD | CPM-4th harmonic | 100 | 11.5 | 25 x 80 Gb/s PAM-4 | 25-100 | w | N/A |
| 2024/[49] | InAs/GaAs QD | CPM-4th harmonic | 100 | 5 | 8 x 25 Gb/s NRZ | 23 | w | N/A |
| 2024/[13] | InAs/GaAs QD | CPM-4th harmonic | 100 | 5 | 8 x 100 Gb/s PAM-4 | 23 | w/o | N/A |
| 2024/[50] | InAs/GaAs QD | CPM-4th harmonic | 100 | 4.8 | N/A | 25-80 | w | 38 |
| 2024/[14] | InAs/GaAs QD | Two-Section | 100 | 13.3 | 24 x 106 Gb/s PAM-4 | 25 | w | N/A |
| This Work | InAs/GaAs QD | CPM–2nd harmonic | 100.6 | 14.319 | 26 x 128 Gb/s PAM-4 @25 °C. 22 x 128 Gb/s PAM-4 @85 °C | 25-140 | w/o | 144 |

In conclusion, this study presents a groundbreaking advancement in 100 GHz QD-MLL technology through the implementation of a co-doping strategy during QD material growth, integrated within CPML architecture. The proposed device achieves record-setting performance for 100 GHz QD comb lasers, outperforming previously reported 100 GHz QD-MLL systems in key metrics. Specifically, the device delivers the widest reported 3 dB bandwidth of 14.312 nm at 25 °C, supporting 26 comb lines each modulated at 64 Gbaud PAM-4, yielding a total throughput exceeding 3.238 Tb/s. Stable mode-locking is maintained up to 140 °C, with 9 tones preserved within the 3 dB bandwidth even at 125 °C. Most impressively, the QD-MLL comb laser sustains 22 channels at 85 °C, enabling a total transmission capacity of 2.816 Tb/s at elevated temperature, demonstrating unprecedented high-temperature performance. The corresponding power consumption of the laser to transmit per bit is 0.394 pJ/bit at 25 °C and 0.532 pJ/bit at 85 °C, respectively. Furthermore, long-term reliability is confirmed through accelerated aging tests, with the MTTF extrapolated to approximately 207 years under continuous operation at 85 °C. The results indicate that the co-doped QD-MLL comb laser offers significant potential to sustain multi-channel modulation at elevated temperatures for high-speed optical interconnects, meeting the stringent demands of future OIO architecture that prioritize energy efficiency, scalability, and thermal robustness. By bridging laboratory-scale characterization with real-world operational requirements, this work represents a key milestone toward the development of high-density, thermally resilient optical interconnects capable of meeting the escalating bandwidth and thermal management demands of next-generation terabit-scale technologies.

## Materials and methods

### Material Growth

The epitaxial sequence commenced with a 300 nm heavily n-type doped GaAs contact layer, followed by a 1.4 μm n-type doped AlGaAs cladding layer. A 60 nm undoped GaAs waveguide layer was subsequently deposited. The active region comprised eight layers of InAs/InGaAs dot-in-a-well (DWELL) structures. Each InAs QD layer was embedded within a 4.5 nm InGaAs quantum well (QW), separated by a 50 nm GaAs spacer layer. Silicon (Si) dopants were introduced during QD nucleation to achieve an electron density of 0.6 electrons per dot (e/dot). Simultaneously, a 10 nm GaAs within each spacer layer between dot layers was p-type modulation doped with a hole density of 10 holes per dot (h/dot). The epitaxial structure was then completed with a p-type doped AlGaAs cladding layer and a heavily p-type doped GaAs contact layer.

### Device Fabrication and Package

Fabrication of laser devices followed standard FP cavity laser fabrication procedures with an additional wet etch step to establish electrical isolation between the gain section and the SA sections. After completing the fabrication process, the wafer was mechanically thinned to a thickness of 120 μm to facilitate subsequent cleaving. The rear and front facets of the comb lasers were coated with 95% high-reflection and 30% low-reflection films, respectively. Then, the laser devices were mounted p-side up on a chip-on-carrier (CoC) without any specialized thermal management and electrically interfaced via gold wire bonds for characterization. The laser output was directly coupled into a tapered lensed single-mode fiber (SMF) to maximize optical coupling efficiency. Finally, the CoC was integrated into a compact 14-pin butterfly package, fitted with a standard Peltier cooler and a thermistor for conducting further experiments.

### Device characterizations

Two identical source meters (Keithley 2450) were used to apply continuous-wave injection current and reverse-bias voltage independently to the separate sections of the MLL. The Thorlabs laser driver CLD1015 was utilized as a TEC controller to enable a stable operating wavelength. For the L–I curve characterization, output power was measured using a power meter (Newport 1938-R) equipped with an integrating sphere-based fiber optic measurement head (Newport 819D-IG-2-CAL2). Optical spectra were recorded using two types of optical spectrum analyzer (OSA). Mapping results were obtained using the OSA (Yokogawa AQ6370D) with a minimum resolution of 0.02 nm. At the same time, the fundamental repetition rate was determined using the ultra-high-resolution OSA (APEX AP2087A), offering a resolution of 0.04 pm. The pulse duration of the MLL was measured using an autocorrelator (APE PulseCheck). Prior to this measurement, the laser output was polarization-adjusted using a polarization controller. RIN measurements of individual tones and the entire optical spectrum were performed using two noise measurement systems: an OEwaves 4000 noise characterization analyzer for the 0-100 MHz range, and a 40 GHz RIN measurement system (SYCATUS, A0010A-040) for the 100 MHz-20 GHz range. For the measurement of individual tones, a tunable bandpass filter (EXFO, XTM-50) and an O-band praseodymium-doped fiber amplifier (PDFA) (FiberLabs Inc., AMP-FL8611-OB-16) were employed to compensate for the coupling and insertion loss. Meanwhile, to avoid overloading the noise analyzers with excessive power, a variable optical attenuator (VOA) was placed after the PDFA to regulate the received power level to approximately –7 dBm.

## Transmission setup

The selected Tone 1 through Tone 6 was filtered out by a tunable bandpass filter (OBPF) (EXFO XTM-50), followed by a PDFA (FiberLabs Inc. AMP-FL8611-OB-16) to preamplifier the optical power to +12 dBm. Then, the amplified optical carrier signal is launched into a 65 GHz lithium niobate Mach–Zehnder modulator (MZM) (Eospace, AX-0MVS-65-PFA-PFA-STU1220) for data modulation. Prior to the MZM, a polarization controller (PC) is used to determine the polarization direction of the input lightwave and enable maximum signal intensity. A pseudo-random bit sequence (PRBS) with PAM-4 format is generated offline in MATLAB and loaded to a 120 GSa/s arbitrary waveform generator (AWG) (Keysight M8194A). The generated electrical signal is then amplified by broadband RF amplifier (SHF S807C, 3-dB bandwidth of 55 GHz) and used to drive the modulator. The modulated optical signal is transmitted in a back-to-back (B2B) configuration. At the receiver side, a variable optical attenuator (VOA) is used to control the received optical power at a 50 GHz photodetector (Finisar XPDV2320R). The detected electrical signal is subsequently amplified by another RF amplifier and digitized by a real-time digital storage oscilloscope (DSO) (Keysight UXR0594A) with a 256 GSa/s sampling rate. Finally, offline digital signal processing (DSP) is applied for signal recovery and BER calculation. For optical eye diagram characterization, the received optical signal was directly measured after the modulator using a sampling oscilloscope (Keysight DCA-M N1092A).

## MTTF analysis

A sub-linear model is used to fit the threshold behavior over time:

$$I_{th}(t) = I_{th}(0)(1 + at^m),$$

$$MTTF = \left(\frac{1}{a}\right)^{\frac{1}{m}}$$

Where $I_{th}(t)$ is the threshold as a function of aging time, $I_{th}(0)$ is the initial threshold current, which is 18 mA. "t" is the aging time, "a" and "m" are constants that need to be determined through numerical fitting.

## Acknowledgement

This work is supported by the National Key Research and Development Program of China (2023YFB2805900) and the National Natural Science Foundation of China (Grant No. 62474171, U21A20454, U23A20356, 62235017). The authors would like to express their appreciation to Dr Zichuan Zhou and Dr Jiajian Chen for their valuable and constructive discussions. The authors are also grateful to the staff members of the National Information Optoelectronics Innovation Center for their continuous technical support.

## Author contributions

S. Pan, V. Cao, X. Xiao, and S. Chen conceived the idea and designed the experiments. J. Yang did the material growth, and C. Zhao performed the material characterization. Device packaging was carried out by J. Yan, and Y. Feng conducted device characterization. S. Pan, V. Cao, and D. Wu executed the experiments. Data curation and formal analysis were performed by S. Pan, V. Cao, and S. Chen. S. Pan drafted the manuscript with input from all authors. X. Xiao and S. Chen initiated the collaboration and provided project supervision. All authors reviewed and approved the final manuscript.

## Data availability

The experimental data that supports the findings of this work is available upon reasonable request from the corresponding authors.

## Conflict of interest

The authors declare no competing interests.